\documentclass{article}

\usepackage{fancyhdr}
\usepackage{amsmath}
\usepackage{arxiv}
\newcommand{\undertitle}[1]{}
\newcommand{\headeright}{}
\usepackage[utf8]{inputenc} 
\usepackage[T1]{fontenc}    
\usepackage{hyperref}       
\usepackage{url}            
\usepackage{booktabs}       
\usepackage{amsfonts}       
\usepackage{nicefrac}       
\usepackage{microtype}      
\usepackage{lipsum}        
\usepackage{graphicx}
\usepackage{natbib}
\usepackage{doi}
\usepackage{float}
\usepackage{graphicx}
\usepackage{subcaption}
\usepackage{parskip} 

\title{Incentivised Orchestrated Training Architecture
(IOTA): A Technical Primer for Release}
\hypersetup{
  pdftitle={IOTA: A Technical Primer},
  pdfauthor={Felix Quinque, Alan Aboudib, Szymon Fonau, Rodrigo Lopez Portillo Alcocer, Brian McCrindle, Steffen Cruz},
  pdfkeywords={Decentralized Training, Bittensor, IOTA, SWARM, LLMs},
}

\author{%
  Felix~Quinque, Alan~Aboudib, Szymon~Fonau, Rodrigo~Lopez~Portillo~Alcocer,
  Brian~McCrindle, Steffen~Cruz\\[0.4em]
  Macrocosmos~AI%
  \thanks{All authors may be contacted at
          \texttt{\{firstname\}@macrocosmos.ai}.}%
  \thanks{\textcopyright{} 2025 Macrocosmos AI. All rights reserved.}%
}

\date{}

\begin{document}
\maketitle

\begin{abstract}
In August 2024, Bittensor's Subnet 9 (SN9) demonstrated that a distributed network of incentivized, permissionless actors could each pretrain large language models (LLMs) ranging from 700 million to 14 billion parameters, while surpassing established baselines \citep{macrocosmos-whitepaper}. While that work validated blockchain-based decentralized pretraining as viable, it contained core issues: (i) every miner had to fit an entire model locally, and (ii) “winner-takes-all” rewards encouraged model hoarding.
\vspace{5pt}

Here we introduce \textbf{\textit{IOTA (Incentivized Orchestrated Training Architecture)}}, an architecture that addresses these limitations by transforming SN9's previously isolated competitors into a single cooperating unit that can scale arbitrarily while still rewarding each contributor fairly. IOTA is a data- and pipeline-parallel training algorithm designed to operate on a network of heterogeneous, unreliable devices in adversarial and trustless environments. The result is a permissionless system that (1) is capable of pretraining frontier-scale models without per-node GPU bloat, and (2) tolerates unreliable devices and (3) aligns participants through transparent token economics. 
\vspace{5pt}

Below, we present the key pieces of work explored in the process of creating IOTA. \textbf{We note these are a series of preliminary results, to be validated in the production release}:

\begin{itemize}
  \item \textbf{Data- and Pipeline-parallel SWARM architecture} – An orchestrator distributes model layers across heterogeneous miners and streams activations between them, enabling model sizes to scale with the number of participants rather than being constrained by the VRAM of a single machine.

  \item \textbf{Granular, continuous incentives} – Validators continually measure each miner's contribution; token emissions are proportional to the work done by each node, rather than the previously utilized winner-takes-all incentive landscape in SN9.

  \item \textbf{Activation compression} - We explore compression techniques via model-bottlenecks to cut communication bandwidths of activations by up to 128$\times$, vastly improving training speed.

  \item \textbf{Butterfly All-Reduce} – Miners average disjoint parameter slices in $\mathcal{O}(1)$ bandwidth, offering linear scalability, redundancy and built-in collusion detection.

  \item \textit{\textbf{CLASP (Contribution Loss Assessment via Sampling of Pathways)}}: A fair attribution scheme assigns credit to miners proportional to their marginal utility and detects exploits, even when contributions are interdependent across the pipeline.
\end{itemize}

\end{abstract}

\newpage

\section{The Landscape of Distributed Pretraining}

The decade of centralized training and algorithm optimization since the AlexNet (\cite{krizhevsky2012imagenet}) moment in deep learning has continued to reinforce what is commonly referred to as \emph{The Bitter Lesson} (\cite{sutton2019bitter}):

\begin{quote}\centering
“General methods that leverage computation are ultimately the most effective.”
\end{quote}

Recent years have seen an explosion in the scale of pretrained models. Particularly in NLP, frontier models have been trained which exceed 1 trillion parameters.  At such a scale, a single model can no longer fit into the memory of one GPU and must be partitioned across many devices\@. Currently, training such models demands intensive high-bandwidth communication between devices and assume reliable, low-latency interconnect, making training feasible only in tightly controlled data-centre environments.  The requisite infrastructure is also notoriously expensive, available only to a few organizations.  This centralization of compute not only raises the financial barrier to entry, but also limits who can experiment and iterate at the cutting edge of model development.

These realities motivate the search for distributed alternatives to centralized pretraining.  Researchers have begun exploring more cost-efficient setups that leverage dispersed resources: for example, renting fleets of cheap pre-emptible cloud instances or pooling volunteer computing power.  Such approaches promise to democratize access by tapping into a “cluster-of-the-whole-internet” in place of a single mega-cluster.  Yet running large-scale training on unreliable, heterogeneous networks presents new challenges.  Traditional data parallelism (DP) and model/pipeline parallelism (MP/PP)—the backbone of today’s LLM training—each face significant trade-offs in decentralized settings.

\emph{Data parallelism} replicates the complete parameter set on each worker, partitions the training corpus, and performs synchronous gradient averaging after every step.  This strategy is implementation-friendly and resilient to slow or failing participants because individual mini-batches can be processed independently.  Communication overhead can be mitigated through high-ratio gradient compression—reductions approaching 800x have been reported without measurable loss in accuracy (\cite{aji2017sparse}\cite{peng2024distro}).  Recent decentralized demonstrations, notably Prime Intellect’s 10 B-parameter \textsc{Intellect-1}(\cite{prime2024intellect} and Templar’s trustless 1 B-parameter \textsc{Templar-I} (\cite{templar2024templar}), confirm that compressed data parallelism can converge on heterogeneous, volunteer GPU clusters.  The principal drawback remains memory footprint: every participant must accommodate the full model and its optimiser states.  Consequently, large-scale DP presupposes access to multi-GPU servers (e.g.\ 8 $\times$ H100), limiting its suitability for broad, permissionless participation.

\emph{Model parallelism} leverages the sequential multi-layer structure of the network so each worker stores only a slice of the weights, allowing models that exceed single-device memory.  Two variants dominate: \textit{tensor parallelism}, which divides computation within each layer but incurs costly all-to-all exchanges after every layer, and \textit{pipeline parallelism}, which assigns contiguous layer blocks to different devices and streams activations forward (and gradients backward).  While pipeline parallelism reduces per-layer traffic, both schemes presuppose reliable, high-bandwidth links; any straggler can stall the pipeline, making conventional MP/PP ill-suited to open, heterogeneous networks.


As such, prior distributed training strategies have faced three fundamental limitations outside centralized clusters: (a) memory constraints if every participant must load the full model (the DP approach); (b) communication bottlenecks and failure sensitivity when splitting models across participants (the MP/PP approach); and (c) without effective incentives, malicious participants can disrupt the delicate process of training AI models.

\emph{SWARM Parallelism} \cite{ryabinin2023swarm} offers a novel alternative by addressing limitations (a) and (b). SWARM is a model-parallel training algorithm explicitly designed for “swarms” of unreliable, heterogeneous machines.  It extends pipeline parallelism with added resilience and adaptivity: instead of a fixed pipeline that fails if one node drops, SWARM dynamically creates randomized routes through the network and reconfigures them on the fly in response to faults or stragglers.  At a high level, the system prioritizes faster and more stable peers for critical pipeline stages and periodically redistributes the work as devices join or leave.  This stochastically wired, fault-tolerant approach reduces the impact of slow or lost participants and represents a viable path for model-parallel pretraining on the kinds of unreliable, globally distributed systems that were previously written off as too slow or flaky.

Meanwhile, addressing limitation (c) has been the focus of the \textsc{Bittensor} network, a framework that introduced a blockchain-based economic layer for generalized incentivization \cite{bittensor}.  Bittensor’s SN9, launched as an experimental subnet for LLM pretraining, has achieved noteworthy results: \textit{decentralized miners collectively developed models up to 14 billion parameters that outperformed comparable industry baselines} (OpenAI GPT-2 Large and Falcon-7B) on perplexity benchmarks \cite{macrocosmos-whitepaper}.  Bittensor’s innovations in incentive design therefore create a pathway to achieving permissionless, performant systems that can organize unprecedented amounts of compute.

In summary, various solutions attempt to solve key technical hurdles regarding distributed training but lack an incentive model, while others provide economic incentives but have yet to achieve the training performance of a coordinated cluster. \textsc{IOTA} bridges this gap by combining novel techniques that jointly tackle all three limitations.

\begin{itemize}
  \item \textbf{\emph{Incentivized Pipeline Parallelism}}: Section \ref{iota-section} introduces IOTA, a training architecture in which a single large model is partitioned across miners in pipeline-parallel fashion.  Each miner processes a slice of the model (a set of consecutive layers), while training data samples stream through the pipeline in a data-parallel manner.  Crucially, the blockchain-based reward mechanism is redesigned so that all participants in the pipeline are rewarded in proportion to their contribution.  By “spreading” the model across loosely connected participants, IOTA enables continuous training beyond the memory limits of any one machine and welcomes hardware ranging from consumer-grade GPUs to cutting-edge accelerators.
  \item \textbf{\emph{Activation Compression}}: In Section \ref{activation-compression-section}, we introduce a novel “bottleneck” transformer block built on Llama3 that preserves residual pathways while compressing activations (and gradients) by up to 128× in bf16, attempting to matching data-center speeds and maintaining convergence even at extreme compression levels.
  \item \textbf{\emph{Butterfly All-Reduce for Trustless Merging}}: Section \ref{sec:BAR} introduces Butterfly All-Reduce, a collective operation to globally aggregate model updates (such as gradient sums or weight averages) from all participants. IOTA employs butterfly all-reduce to perform data aggregation in a decentralized and verifiable manner.  This serves as a “merge primitive” for assembling contributions from many miners into one global model without relying on any central server and is resilient to a fraction of malicious or dropped participants. 
  \item \textbf{\emph{CLASP}}: To fairly attribute credit in multi-node training, IOTA introduces a Shapley-value-based algorithm called \textit{CLASP: Contribution Loss Assessment via Sampling of Pathways} in Section \ref{shapley-section}. Shapley values, originating from cooperative game theory, quantify each participant’s marginal contribution to the model's improvement. Rather than relying on simplistic metrics (e.g., local accuracy), CLASP evaluates the effect of removing or substituting individual miners on convergence and model quality. This not only promotes honest effort and deters free-riding—since rewards are tied to provable impact—but also serves as a mechanism for detecting adversarial or exploitative behavior. While highly promising, CLASP is not included in the initial release; it remains an active area of research and is intended for integration into the incentive mechanism once the system stabilizes post-launch. Further publications will detail the full design and its empirical foundations.
\end{itemize}

In this report we introduce the technical architecture and innovations behind these four components in greater depth, together with proposals for initial implementation and avenues for further research.

\section{IOTA Architecture}
\label{iota-section}
\begin{figure}[H]
    \centering
    \includegraphics[width=0.6\linewidth]{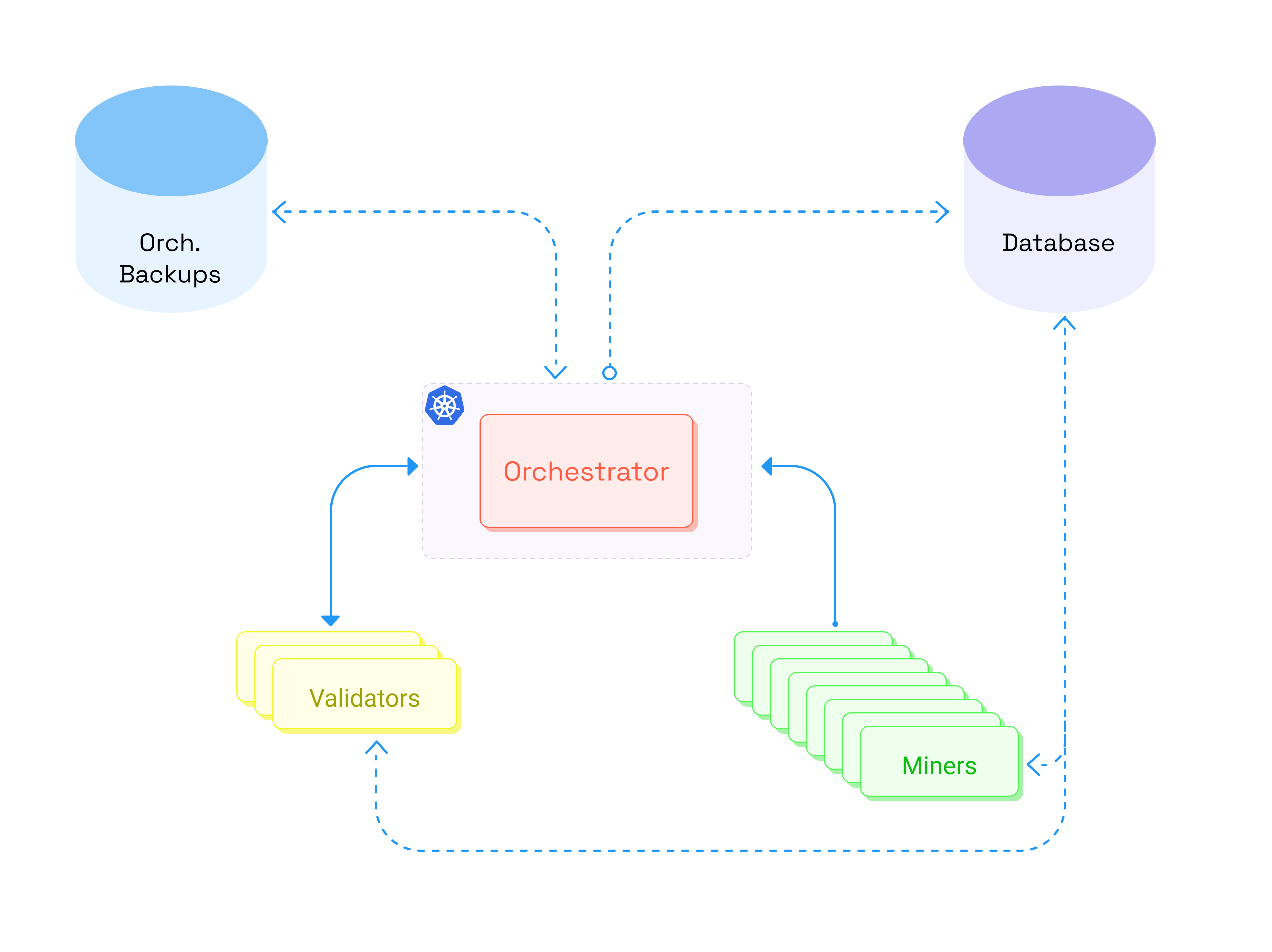}
    \caption{Overall system architecture. The orchestrator facilitates the training process by triggering miners to work on specific layers of the model, further triggering when validation should occur based on the progress of the miners }
    \label{fig:IOTA}
\end{figure}

IOTA is structured around three core roles: the Orchestrator, Miners, and Validators. The simplified design of the system is illustrated in Figure \ref{fig:IOTA}. Rather than adopting a fully peer-to-peer topology, IOTA follows a hub-and-spoke architecture centered around the Orchestrator. This design choice ensures global visibility and enables comprehensive monitoring of all interactions between participants, which is critical for enforcing incentives, auditing behavior, and maintaining system integrity.

\begin{figure}[H]
    \centering
    \includegraphics[width=1\linewidth]{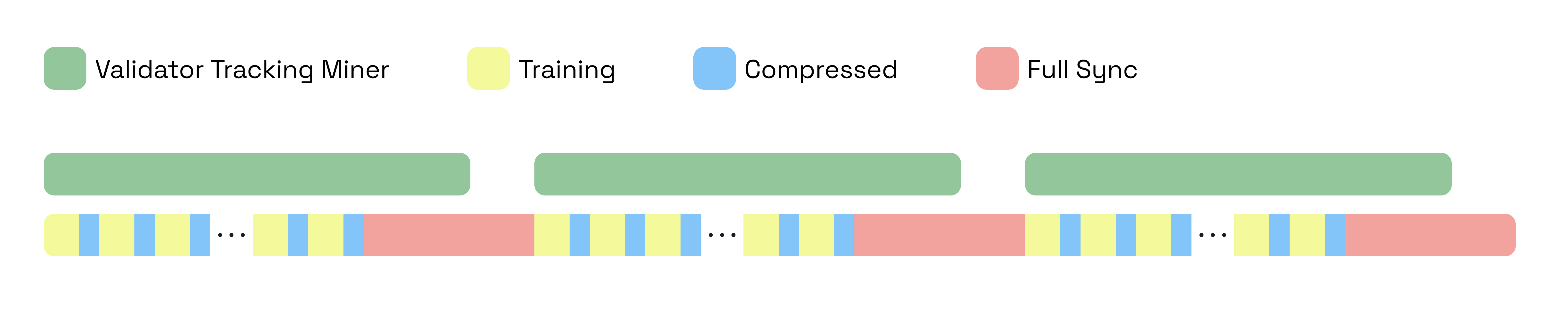}
    \caption{Timeline of validator-miner tracking, training, compressed sharing, and syncing blocks. Each stage of the learning process is triggered by the orchestrator to facilitate training.}
    \label{fig:timeline}
\end{figure}

This architecture allows a system-level orchestrator to manage how participants on the network will operate at different stages of the training process. All data that is created and handled by these three entities is pushed to a globally accessible database, making it easy to trace the movement of information. Figure \ref{fig:timeline} illustrates the temporal relationship between model training, validator-miner tracking, data sharing, and model sync, which are all triggered by the orchestrator. 

\begin{enumerate}
    \item In the \textit{training stage}, miners process forwards and backwards activations. First layer miners read from the dataset, middle layer miners each then apply their layer to the activations until the last layer miners compute the loss. Finally, the whole process is reversed to compute the gradients for the entire LLM.
    \item During \textit{compressed sharing}, miners upload a highly compressed version of their weights and optimizer state to be shared within their layer. Since the validator has processed identical activations, both the validator and miner should have identical local states. There may be small deviations (e.g. if either miner or validator skipped an activation), but they show very high similarity.
    \item After $n$ training and compressed sharing stages, a \textit{full synchronization} is required. During this stage, miners share their full uncompressed weights and optimizer states within their layer. This stage allows new miners to join in by copying the existing miners' states, as well as validators to replicate a miners' state for validation.
    \item In the \textit{validation stage}, a validator chooses a miner at random and reproduces its actions. It will then confirm that its own results are aligned with the miners results, and assign a score as described in Section \ref{incentivation-section}. 
\end{enumerate}

The stage between two full synchronization steps (and therefore also the length of a validation stage) will be referred to as an "epoch".

\subsection{Orchestrator}
The orchestrator's primary responsibility is to monitor the training progress of each miner over all discrete layers and initiate weight‐merging events accordingly. Given the heterogeneous nature of miner hardware and their unreliability, it is impractical to wait for all miners to complete an equal number of batches $B$. Instead, we define a minimum batch threshold, \(B_{\min}\), that a miner must complete for its contribution to be considered in the merging process. Once at least a specified fraction of miners have trained for at least \(B_{\min}\) batches, the orchestrator prompts all qualifying miners to upload their weights.

This strategy ensures robustness to stragglers and allows us to define the \emph{effective batch size} $B_\text{eff}$:

\[
    B_{\mathrm{eff}}
    \;=\;
    \sum_{m=1}^{M} \begin{cases}
        B_m & \text{if} \ B_m \geq B_{min} \\
        0 & \text{if} \ B_m < B_{min}
    \end{cases},
\]

where \(M\) is the total number of miners and \(B_m\) is the number of batches completed by miner~\(m\).

This mechanism draws inspiration from centralized training practices—where \(B_{\mathrm{eff}}\) mimics the behavior of global batch sizes in typical LLM training—but in the decentralized setting it is coupled with DiLoCo \citep{douillard2024dilocodistributedlowcommunicationtraining}, which enables miners to perform local optimization steps independently before synchronization. DiLoCo is particularly well suited for this paradigm, as it:

\begin{itemize}
  \item Embraces partial participation from miners,
  \item Supports asynchronous and layer‐wise updates, and
  \item Reduces communication overhead by focusing on the most informative coordinate updates locally.
\end{itemize}

\subsection{Miners}
Miners may register to the subnetwork at any time. Upon registration, the orchestrator assigns each miner a model layer to train. The miner will wait until the next full synchronization period to start actively participating. During the full synchronisation, it will update its weights and optimizer states to align with the rest of the network, and can then proceed to processing forward and backward activations in the training stage.

\begin{figure}[htbp]
  \centering
  \begin{subfigure}[b]{0.46\linewidth}
    \centering
    \includegraphics[width=\linewidth]{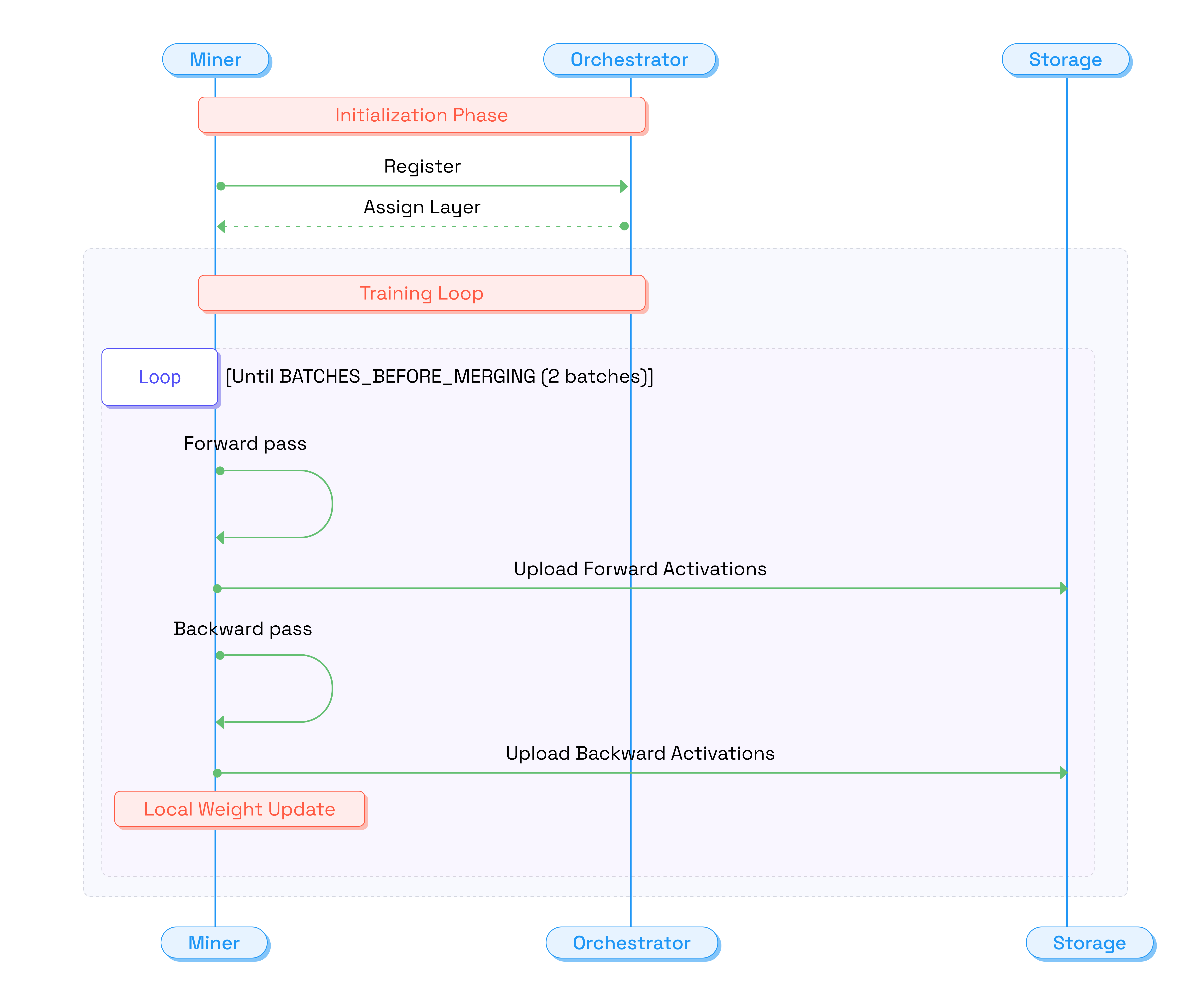}
    \caption{Illustrates the relationship between a miner and the orchestrator. Training occurs for \texttt{BATCHES\_BEFORE\_MERGING} (also known as $B_{min}$) before weight merging is triggered.}
    \label{fig:training_loop}
  \end{subfigure}
  \hfill
  \begin{subfigure}[b]{0.52\linewidth}
    \centering
    \includegraphics[width=\linewidth]{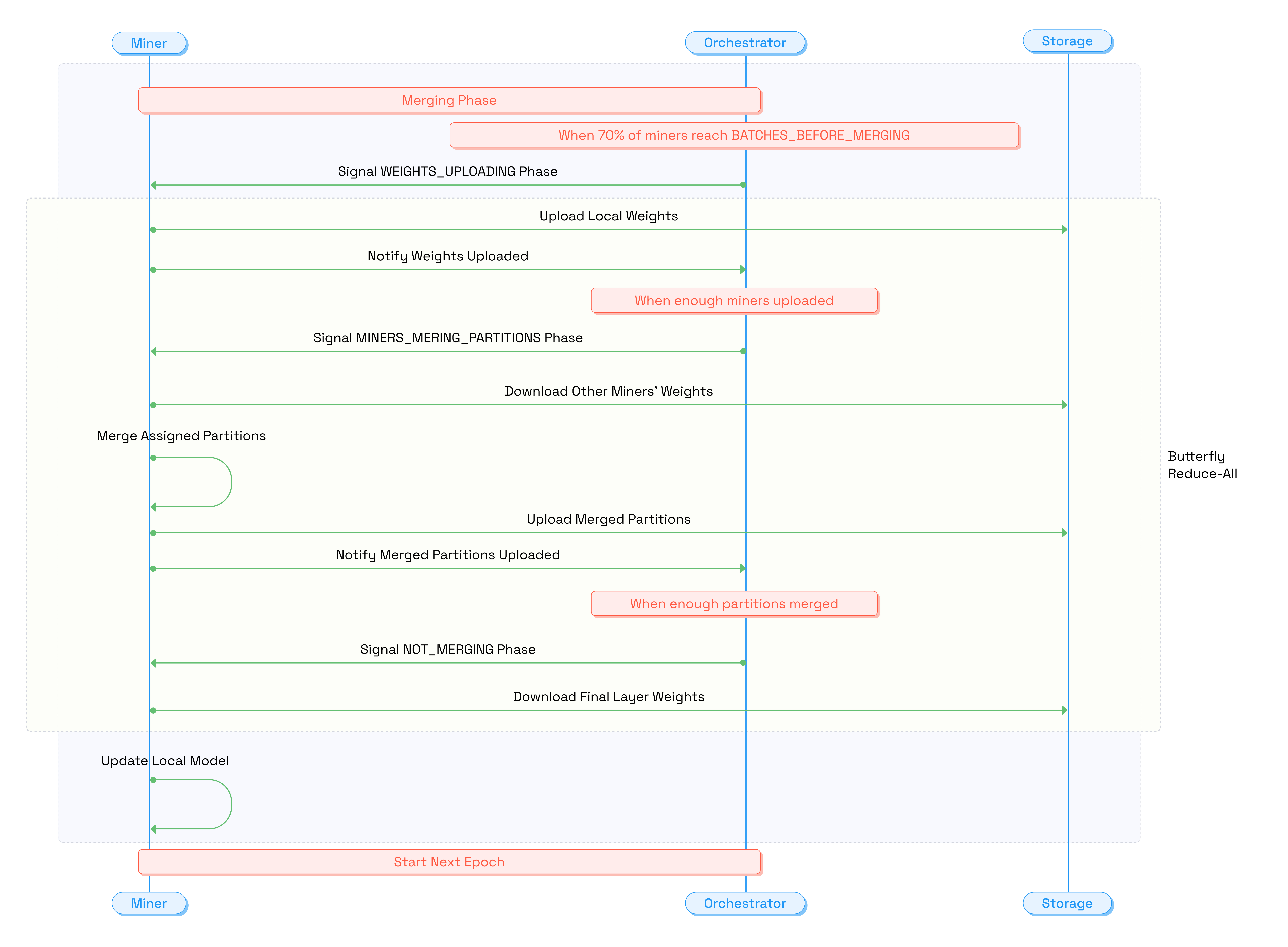}
    \caption{Illustrates the relationship between the miner and the orchestrator during model merging and update. Outlined is where the Butterfly All-Reduce methodology is applied, discussed further in Section~\ref{sec:BAR}.}
    \label{fig:merge_loop}
  \end{subfigure}
  \caption{%
    The two primary stages of the IOTA network.%
  }
  \label{fig:combined_loops}
\end{figure}


During the training loop, as illustrated in Figure \ref{fig:training_loop}, the miner performs forward and backward passes while uploading activations to a dedicated storage bucket. In the forward pass, miners receive activations from the previous layer, compute transformed outputs, and propagate them downstream. In the backward pass, they consume gradients, compute local weight updates, and send gradients upstream. The number of forward/backward passes per loop is controlled by the orchestrator hyperparameter \texttt{BATCHES\_BEFORE\_MERGING} (i.e.\ \(B_{\min}\)). Once \(B_{\min}\) batches are complete, the orchestrator triggers the merging stage, illustrated in Figure \ref{fig:merge_loop}.


Miners periodically synchronize their gradients and less frequently weights \& optimizer states with peers assigned to the same layer, contributing to a global merging process orchestrated by the central coordinator. Miners in the first layer also handle data ingestion and tokenization, while those in the final layer compute the training loss. This division of responsibilities enables a fully parallel, layer‐wise training pipeline.

\subsection{Validators}
Within the system, validators play a core role in determining if the work completed by the miner was honest. Primarily, the validator relies on computational reproducibility to achieve this validation signal. As the validator is tracking a specific miner, a portion of the miner's training is completely rerun on the validator side. Forward and backwards passes are checked against the submitted miner activations using a cosine similarity. However, there are many complications when it comes to reliable validation, and we explore them in the remainder of the paper. We formalize incentivization in Section \ref{incentivation-section}, and add additional exploratory techniques leveraging Shapley values for anomaly detection and adversarial robustness in Section \ref{shapley-section}.

\section{Incentivization}
\label{incentivation-section}
The design of the incentive landscape for the network participants should consider the trade-offs between optimization and reproducibility, and has significant impact on the dynamics of the system. As discussed above, validation hinges on the validator's ability to reproduce sections of training to a chosen threshold. Given this condition, the design does not give power to the miner to innovate algorithmically at this time.

Validators continuously monitor randomly assigned miners throughout full sync stages to ensure comprehensive oversight. To maximize validation coverage, the initial implementation of IOTA employs the shortest possible monitoring period (utilizing 0 compressed sharing stages), enabling each validator to oversee the maximum number of miners within the network. Importantly, miners are not aware of when they are being monitored, preventing them from selectively behaving correctly only during observed intervals. Upon completion of a validation stage, the mining rewards are calculated based on the total number of backward passes successfully processed, $S_m^n$, where $m$ indexes the miner and $n$ indexes the validation epoch.

The system implements a temporal decay mechanism governed by hyperparameter $\gamma$, which determines the decay time. The weight decay for miner $m$ in epoch $n$ follows a step function -- concretely this means that a miner is assigned a fixed amount of "score" for a time period $\gamma$, after which the score drops to 0:

\[w(t)_m^n = \begin{cases}
1 & \text{if } t \leq t_{\text{decay}} \\
0 & \text{if } t > t_{\text{decay}}
\end{cases}\]

where $t$ is the time since the score was initially assigned.
Therefore, the raw incentive $I$ is the sum over all scores multiplied by their time weighting factor $w(t)_m^n$
\[
I_m = \sum_{n=0}^N S_m^n \cdot w_m^n(t)
\]

where $N$ is the total number of full synchronization steps at that point of time. This simple linear reward structure ensures miners receive fixed compensation per processed activation, eliminating incentives for throughput manipulation or other gaming strategies during non-validation periods. The exact recomputation requirement during validation stages provides additional security against system exploitation. The stability analysis of this incentive landscape, including numerical simulations demonstrating equilibrium properties, is detailed in Appendix \ref{appendixA}.

\section{Activation Compression}
\label{activation-compression-section}

In model parallelism, a neural network is partitioned across several devices. During the forward pass, each device transmits its computed output activations to the subsequent device in the pipeline. Then, in the backward pass, the activation gradients are communicated in the reverse direction. This inter-device communication of activations and their gradients introduces substantial overhead, particularly when conducted over the Internet with typical bandwidths ranging from 50 Mbps to 200 Mbps. In contrast, high-speed interconnects employed within data centers, such as NVLink (offering bandwidths up to 900 GB/s) and InfiniBand (providing speeds reaching 200 Gb/s and beyond), significantly mitigate these communication bottlenecks.

It is generally agreed that achieving transfer times over the Internet comparable to those in data center environments requires compressing activations and their gradients by approximately 100x to 300x, depending on factors such as model scale and architectural complexity\cite{ryabinin2023swarm,peng2024distro}. Prior approaches for compressing these tensors include 8-bit quantization or dimensionality reduction through the insertion of bottleneck layers at communication boundaries, usually between transformer blocks in transformer-based architectures. However, 8-bit quantization offers a limited compression ratio of 4x relative to 32-bit floating-point precision (fp32). Furthermore, integration of bottleneck layers has been shown to induce a more pronounced decline in convergence rates compared to quantization techniques \citep{ryabinin2023swarm}. 

In our experiments, we found that the introduction of bottleneck layers in between transformer blocks leads to a substantial reduction in convergence. We primarily attribute this degradation to the disruption of residual connections rather than to the reduced dimensionality of the activation tensors. Residual connections have been shown to be crucial for unimpeded gradient propagation throughout deep network architectures, such as transformer-based LLMs, making them more prone to the vanishing gradient problem \citep{rothman2021transformers}.

\begin{figure} [H]
    \centering
    \includegraphics[width=0.7\linewidth]{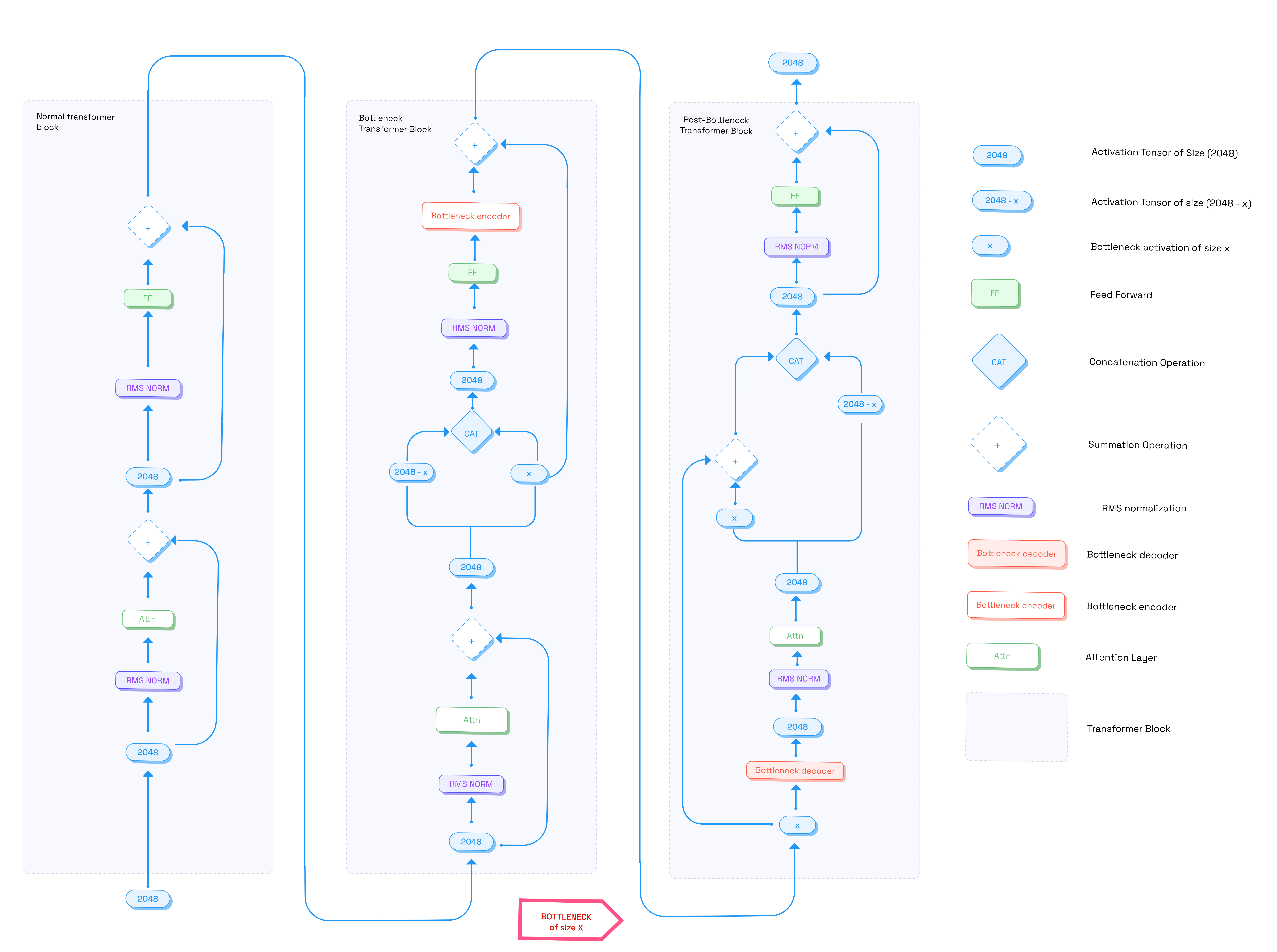}
    \caption{Bottleneck-LLama with uninterrupted residual flow. Residual connections are partially fed into output activations of attention layers within bottleneck and post-bottleneck transformer blocks.}
    \label{fig:bottleneck_llama}
\end{figure}

Illustrated in Figure \ref{fig:bottleneck_llama}, we propose a novel bottleneck unit architecture derived from a modified Llama3 transformer block \citep{grattafiori2024llama3herdmodels}. Preliminary results indicate that this architecture enables significantly higher compression rates with minimal impact on convergence, while preserving gradient flow through residual connections. The figure shows three transformer block types. The vanilla Llama3.2 transformer block (left), the proposed bottleneck transformer block (middle), and the proposed post-bottleneck transformer block (right). Partial residuals flow into and from hidden activation at the output of attention layers. 

\begin{figure}[H]
    \centering
    \includegraphics[width=0.7\linewidth]{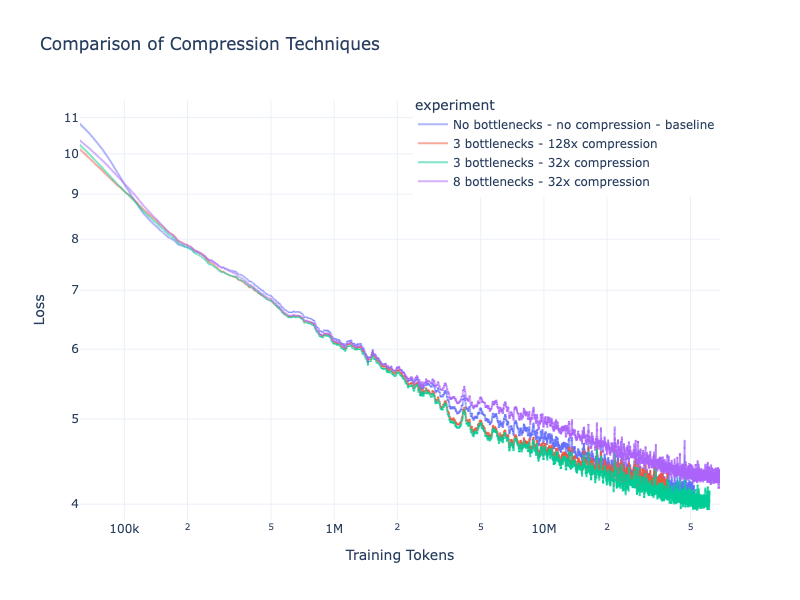}
    \caption{Convergence rate comparison. Early training loss of the baseline Llama3.2-1.5B model compared to our proposed modified Llama3.2-1.5B architecture with different number of bottlenecks and different compression rates.}
    \label{fig:compression_loss}
\end{figure}

As shown in Figure \ref{fig:compression_loss}, our preliminary experiments achieved a 128x symmetrical compression rate for both activations and their gradients, with no significant loss in convergence when training a modified 1.5B parameter three-bottleneck Llama3 model on up to 400 million tokens from the FineWeb dataset \citep{penedo2024the}. Notably, increasing the compression ratio from 32x to 128x resulted in only a slight degradation in convergence. Similar behavior is expected in our ongoing experiments with eight-bottleneck models. While these results are promising, it is imperative to validate them on longer training horizons and with larger datasets—up to 100 billion tokens—to ensure robustness and generalizability.

The compression rates we report are all computed relative to 2048-dimensional, fp32 activation vectors as in the original Llama3 1.5B implementation. All of our compressed activations are computed in bf16 (2x compression). A 32-dimensional bottleneck represents an additional 64x reduction for a total of 128x compression rate.

It should be noted that adding eight bottlenecks to the 16-layer 1.5B Llama3 model is an extreme compression case where 50\% of the transformer blocks are replaced by bottleneck blocks. We expect larger models such as the 80-layer 70B Llama3 model to be more resilient to the introduction of eight bottlenecks, which represents only 10\% of the total number of transformer blocks.

\section{Butterfly All-Reduce}
\label{sec:BAR}
Butterfly All-Reduce \citep{original-butter} \citep{DBLP:journals/corr/abs-2103-03239} is a communication pattern used in distributed computing to efficiently aggregate and distribute data (such as gradients, model weights, and/or optimizer states) across multiple participants. In order to deal with extensive loading times, as well as to reduce validation overhead, we formulate a novel version of butterfly all-reduce below. 

A canonical idea for weight sharing is that in a system of $N$ miners, each miner simply splits their weights into $N$ sections. Each miner is then assigned one section, which it will download from every other miner, merge (e.g. by averaging) and then re-upload. Finally, each miner downloads all merged sections. This na\"ive approach already allows us to share our gradients/weights in $O(1)$ time, but critically does not allow us to easily validate miners or be resistant to unreliable nodes in the network.

Let us begin by creating every possible pairing of miners on a single layer as set P:

\begin{equation}
    P = \{(i,j) | 0 \leq i \leq N - 1\}
\end{equation}

where $P$ has the cardinality $\frac{N(N-1)}{2}$ and therefore grows quadratically with the number of nodes $N$ in the system. From here, we can simply index each element of $P$ in a random order,

\begin{equation}
    f : P \rightarrow \{0,1,2,...,|P|-1\}
\end{equation}

where $f$ is a random function that maps each pairing to a random integer in the range $[0, |P|-1]$. Concretely, in the case of 3 miners, this can create the following mapping: $\{[0: (1,2), 1: (0,2), 2: (0,1)]\}$. As such, each index corresponds to a ‘shard’ of the weights, and each tuple corresponds to a set of two miners. 

We leverage the above foundations to enable significant training speed ups across all stages of communication, described below. 

\begin{figure}
    \centering
    \includegraphics[width=0.7\linewidth]{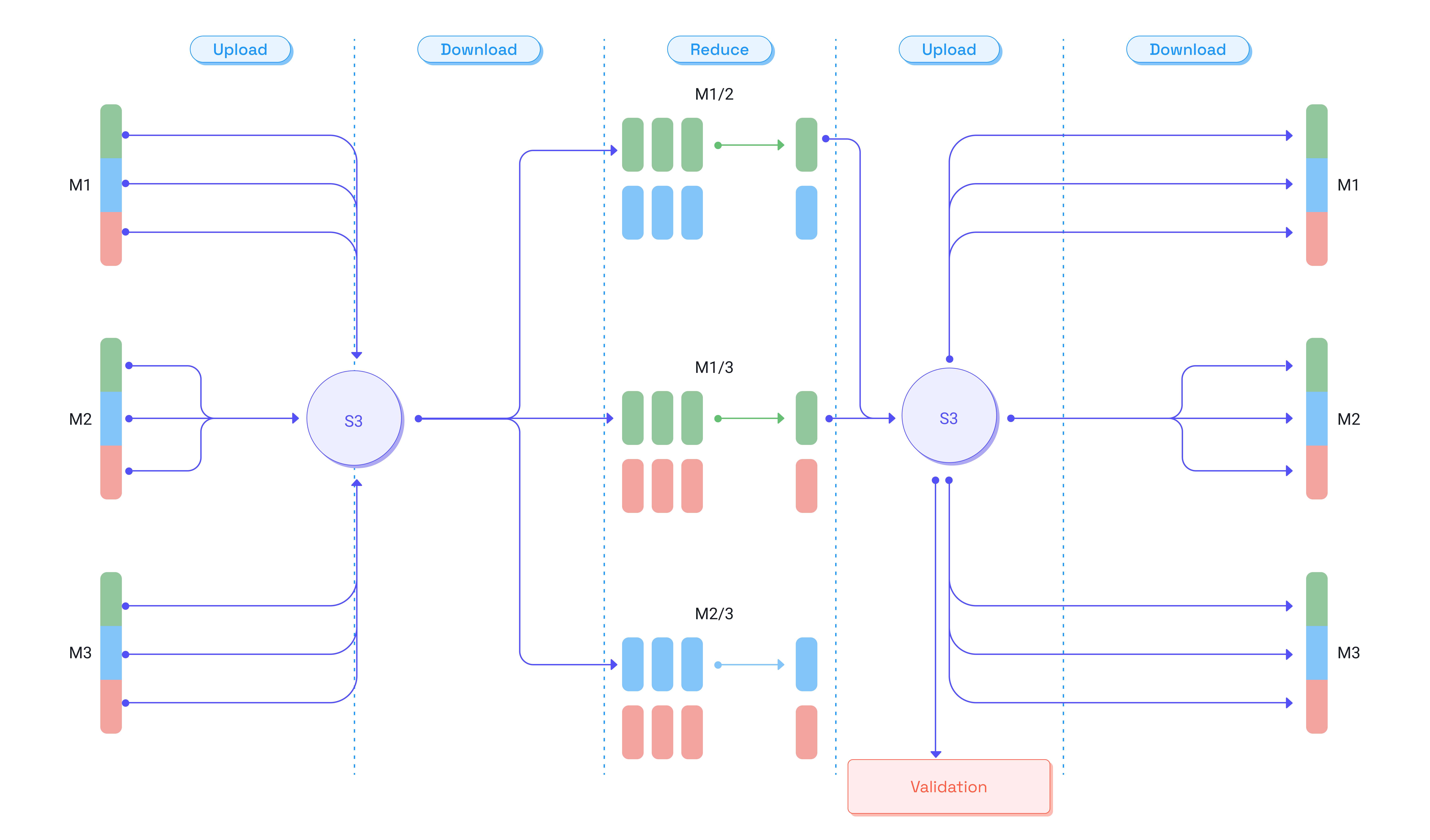}
    \caption{System schematic of Butterfly All Reduce, utilizing an S3 bucket as an intermediate data storage solution. Miners, $M_1$, $M_2$, $M_3$, have their weights sharded into $N = 3$ segments to be uploaded to readable storage. An arbitrary reduction method such as averaging or an outer optimizer step is used to reduce the same shards into a single shard to be redistributed to all all miners for weight-syncing.}
    \label{fig:butterfly}
\end{figure}

\subsection{Weight Upload}
Each miner uploads weights and a corresponding metadata file to enable sharding. The shards consist of equally sized segments of the original tensor, and the metadata maps the starting byte of the saved tensor to an index. In the non-compressed case, this is simply equivalent to $\texttt{bytes\_per\_weight} \cdot \texttt{weight\_index}$ , but once compression is introduced, metadata is required. Figure \ref{fig:butterfly} illustrates the case for 3 miners ($P = 3$), and therefore each miner's uploaded weights (colour-coded for each miner) come with metadata defining their splits into 3 shards.

\subsection{Weight Reduce}
During shard reduction, miners download shards according to the mappings created above. Each miner then averages the shards it has been assigned to and uploads them again to the database. Without loss of generality, we refer to averaging as a simple element-wise geometric mean, but this can be easily extended to more sophisticated averaging techniques (such as outlier-robust averaging \cite{pillutla2022robust, damaskinos2019aggregathor, yin2018byzantine, blanchard2017machine, gorbunov2022secure} or outer optimizer steps).

Importantly, as our mapping ensures that every possible combination of miners is present, it means that each miner ‘shares’ one shard with every single other miner. This means that \textit{every} miners' work is replicated by every other miner, making it trivial to detect cheating miners. While this approach introduces overhead equal to the degree of redundancy that it used, it bolsters the network by adding fault-tolerance, which is highly suitable for a network comprised of unreliable nodes. 

\begin{figure}[H]
  \centering
  \begin{subfigure}[b]{0.48\linewidth}
    \centering
    \includegraphics[width=\linewidth]{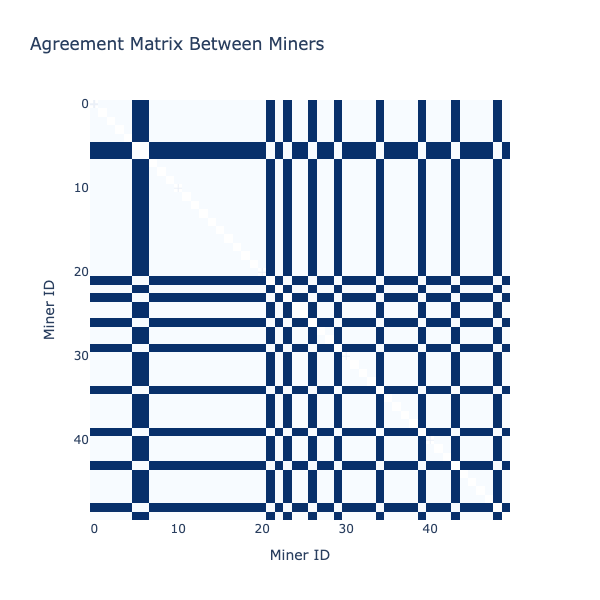}
    \caption{The agreement matrix shows all possible pairings of 50 miners and whether they agree on their submitted weights. All 10 miners with dark lines are deceptive.}
    \label{fig:agreement-rates}
  \end{subfigure}
  \hfill
  \begin{subfigure}[b]{0.48\linewidth}
    \centering
    \includegraphics[width=\linewidth]{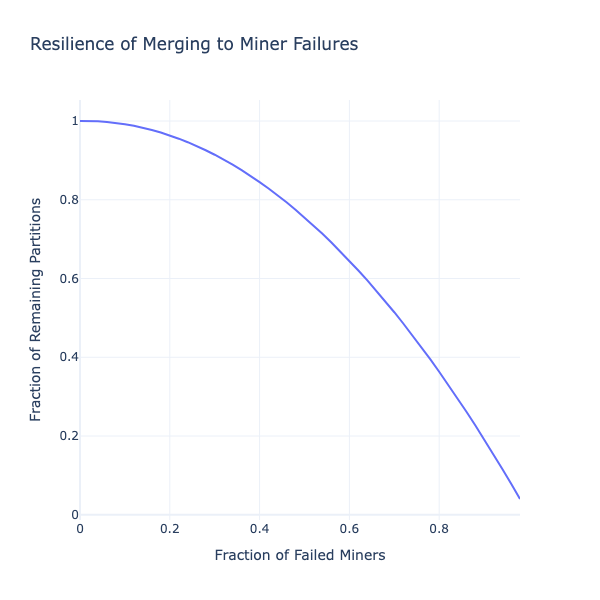}
    \caption{The fraction of remaining averaged weights based on the number of failed miners. Even if 10\% of miners fail, over 95\% of weights are still averaged correctly.}
    \label{fig:resilience}
  \end{subfigure}
  \caption{a) Illustrating the agreement matrix between miners and b) the resilience of miner failures for Butterfly All-Reduce
  }
  \label{fig:merging-fails}
\end{figure}

As miners are not aware of the global splits and only know which sections are assigned directly to them, this technique also prevents collusion between cabalistic miners. To illustrate this point, Figure \ref{fig:agreement-rates} shows the agreement matrix for a system of 50 miners that merge a set of weights using the above technique. It is evident that malicious actors are out of consensus with their peers.

Another beneficial feature is that as miners drop out, we have a high degree of redundancy, allowing us to maintain the majority of weights correctly. Concretely, the number of correct shards can be calculated as 

\[|P_{\text{valid}}| = \binom{N}{2} - \binom{k}{2}\]

where $N$ is the number of miners and $k$ is the number of faulty miners. We can then calculate the percentage of valid shards ($p_{valid}$) as

\[
p_{valid}=1-\frac{\left(N\left(N-1\right)-k\left(k-1\right)\right)}{N\left(N-1\right)}
\]

Figure \ref{fig:merging-fails}(b) shows the resilience of this approach to miner failures by analyzing the fraction of miner weights that are successfully averaged during merging as a function of failure rate in the layer. Given that training is expected to proceed unhindered with around 90\% of the weights successfully merged in a layer, our analysis demonstrates that the system is tolerant to failure rates up to 35\%. Up to failure rates of 10\%, we see virtually no negative impact, as over 99\% of weights are retained.

\subsection{Data Transfer Analysis}
During download, all miners retrieve the merged shards, subsequently reconstructing the full averaged weights. This method is highly data efficient by having $\mathcal{O}(1)$ complexity, regardless of the number of miners in the model. Each miner initially uploads their full weights $W$, followed by downloading $2W$. Afterwards, each miner uploads merged shards of size $\frac{2W}{N_m}$, where $N_m$ is the number of miners in layer $m$. Importantly, the amount of data that must be re-uploaded \textit{decreases} with increasing $N_m$. Finally, each miner downloads a full copy of $W$, meaning each machine must transfer a total of $4W+\frac{2W}{N_m}$. In comparison, a central weight merger would need to transfer $N_m W + 3$, further reinforcing the need for additional compression methods [Section \ref{activation-compression-section}].

\section{CLASP: Contribution Loss Assessment via Sampling of Pathways}
\label{shapley-section}
In data‐ and pipeline‐parallel training runs, bad actors may attempt to “free ride” or even poison the training by producing invalid or corrupted layer activations. This can be catastrophic for training due to the strongly interconnected topology of the training architecture. However, using a relatively straightforward heuristic approach inspired by Shapley values \cite{shapley:book1952}, we present a technique that enables online detection of outliers in the system. Our approach is as follows: samples are routed through the network in a random order (as designated by the orchestrator), and by failing to produce valid activations (either due to omission or adversarial tampering), malicious miners exhibit abnormally high per-sample losses. Over time, as many samples are processed, the orchestrator records both the losses and the “pathways” that samples took through the network. Validators can periodically request these loss‐and‐pathway records and use them to determine the impact on loss of each miner, in a fashion reminiscent of Shapley values. In practice, this can be as simple as calculating the expected loss in an ablation‐like study for each miner, treating each miner as if it were a feature in a dataset.

Fortunately, it is non-trivial for bad actors to accurately impersonate honest peers due to the highly nonlinear dynamics of activations and their interactions throughout the network. The result is that malicious miners are unambiguously flagged because of their consistent association with high losses. With simple scaling techniques (e.g., normalizing by the number of occurrences of a given miner and using z‐scores or similar statistics), this provides a lightweight and reliable means for validators to detect and punish deleterious actions, while also adapting to the evolving loss landscape during training.

This approach rests on two core assumptions: first, that the per‐sample loss (and gradient‐impact) measurements are accurate and tamper‐proof; and second, that miners act without cross‐layer coordination—since they only see black‐box activations from adjacent layers and incur penalties for requesting activations without actually computing them. For this reason, they cannot reliably collude or “window‐shop” to spoof honest behavior. Regarding the first assumption, aggressive spot‐checking of the reported losses in conjunction with redundancy techniques across multiple miners would mitigate the effect of inaccurate outputs. Alternatively, by requiring miners to submit only top-\(k\) compressed logits, validators can recompute exact losses -- ensuring tamper-proof reporting in a bandwidth-efficient, fully decentralized manner that is both practical and secure on Bittensor.

\begin{figure}[H]
  \centering
  \begin{subfigure}[b]{0.48\linewidth}
    \centering
    \includegraphics[width=\linewidth]{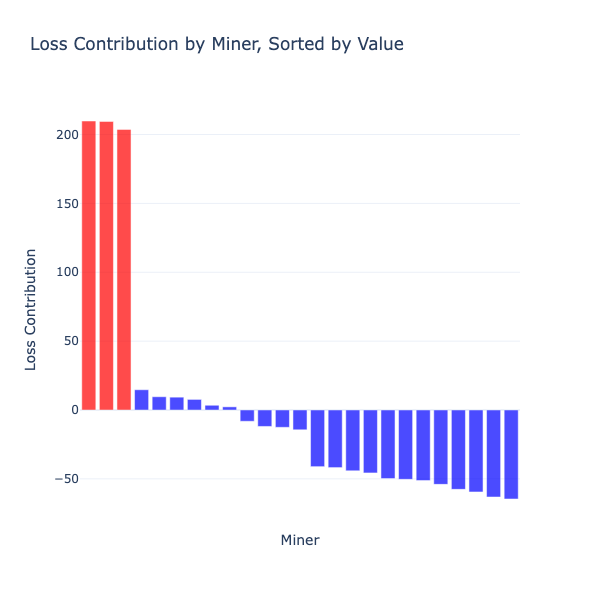}
    \caption{When sorted in descending order of magnitude, the results show that bad actors (red) introduce outsized losses compared to their peers during training.}
    \label{fig:shap-value}
  \end{subfigure}
  \hfill
  \begin{subfigure}[b]{0.48\linewidth}
    \centering
    \includegraphics[width=\linewidth]{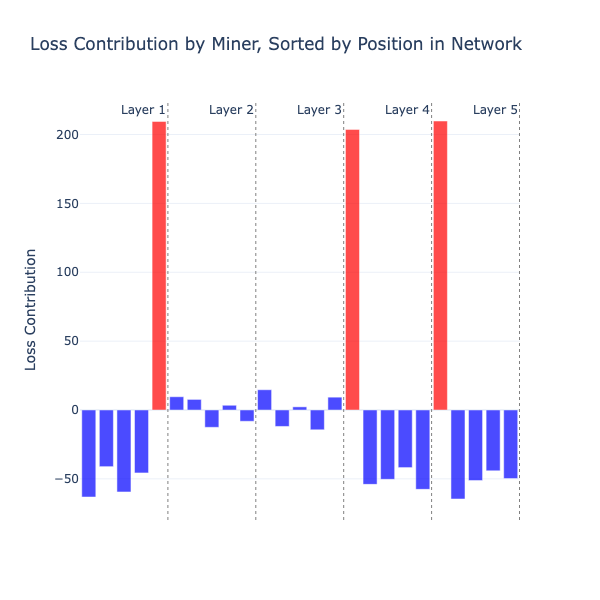}
    \caption{When sorted by network layer, fair miners  in the same layers as bad actors (red) benefit from a commensurate reduction in their loss contribution.}
    \label{fig:shap-position}
  \end{subfigure}
  \caption{%
    Experiment results for loss contributions using CLASP.%
  }
  \label{fig:shapley-plots}
\end{figure}

We present a toy model of this algorithm which simplifies the network and represents the propagation of activations as a simple stochastic sampling process. The model assumes a 5 layer system with 5 miners per layer. Our toy model assumes that our LLM trains with a normally distributed loss of $4.5$, where each batch loss has a standard deviation of $0.2$, which are typical values for models during early training. If a malicious miner is part of the path, we assume that both loss and standard deviation increase by 10\% -- in reality this increase should be significantly more pronounced. Figure \ref{fig:shap-value} shows that malicious actors (red) produce very high outlier scores in the present framework due to their disproportionate impact on loss. Interestingly, Figure \ref{fig:shap-position} also indicates that Shapley values of fair miners in the same layer as bad actors are reduced as a result of an intrinsic balancing process in the technique. This serves to further enhance the sensitivity to outliers. As this is an area of active research at Macrocosmos, a more complete demonstration accompanied by a fully reproducible code will be provided in a subsequent report.

\bibliographystyle{plain}

\section{Summary}

\textbf{Architectural advance.}  
IOTA unifies heterogeneous miners into a coordinated training fabric through SWARM data– and pipeline-parallelism, 128$\times$ activation compression, and an \(\mathcal{O}(1)\) Butterfly All-Reduce.  In early experiments the scheme sustains near-baseline convergence on a 1.5 B-parameter model while freeing each participant from full-model VRAM limits.

\textbf{Economic alignment.}  
Layer-level rewards are distributed continuously and audited by validators, using the CLASP attribution rule inspired by Shapley values. This approach replaces traditional winner-takes-all incentives with more granular compensation. Simulations indicate the emergence of stable equilibria with synchronization windows under an hour, enabling timely feedback without incurring excessive bandwidth costs.

\textbf{Path to production.}  
These findings are preliminary but promising.  On \textbf{2 June 2025} the stack will graduate to Bittensor main-net, where reliability, throughput and incentive dynamics will be tested at internet scale.  We will follow up with a public development roadmap and further work that details the algorithms, fault-tolerance guarantees, and scalability results.

\newpage

\bibliography{references}  

\newpage

\appendix
\section{Appendix A: Numerical Simulation of Incentive Stability}
\label{appendixA}
\subsection{Throughput Efficiency and vTrust}

In order to facilitate the fastest possible training, we would like to minimize the number of full synchronization steps as much as possible due to the large amounts of data that must be transferred. 
The longer the time between synchronizations, the longer a validator will need to process a given number of miners, which in turn will lead to unstable weights. The fewer scores a miner carries at any given time, the more unstable the incentives will be.
\\
The number of scores a miner is expected to have at a given time is simply
\[N_\text{scores}=\frac{\gamma}{T_{s}}\]
where $\gamma$ is the weight decay rate, and $T_s$ is the time between full synchronization steps.

Clearly, by increasing the weight decay $\gamma$ we can increase the number of weights per miner and therefore the weight stability. Unfortunately high values of $\gamma$ also require a longer immunity period for miners and hence slow down the subnetwork as well as delaying the rate of feedback for miners. That means that it allows bad miners to stay on the network for longer and slows down iteration speed.

\begin{figure}[H]
    \centering
    \includegraphics[width=0.5\linewidth]{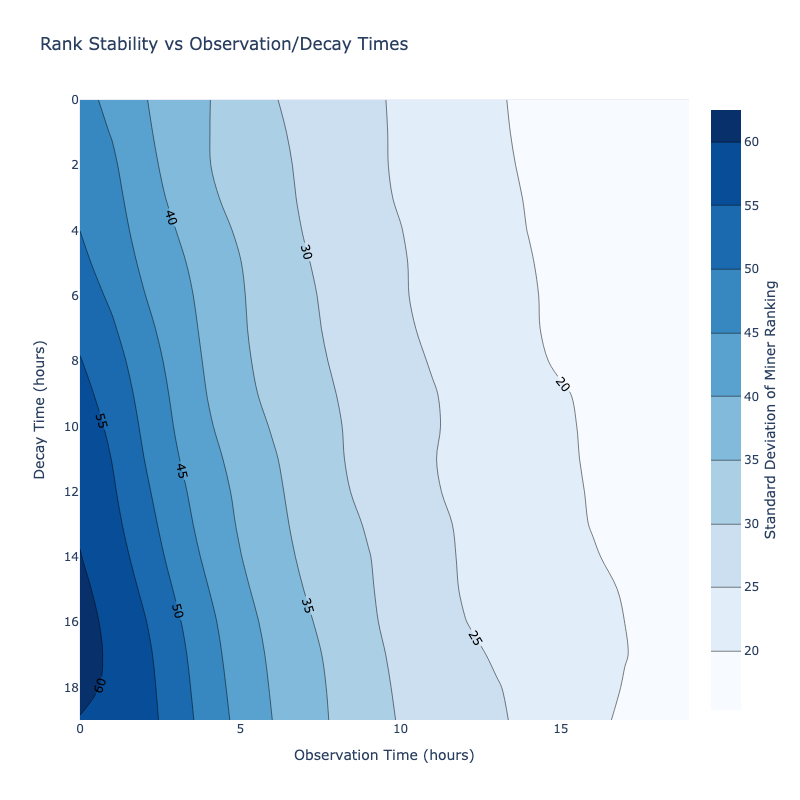}
    \caption{Incentive Stability in Dependence of Monitoring time and Weight Decay Period}
    \label{fig:incentive-stability}
\end{figure}

As Figure \ref{fig:incentive-stability} shows, numerical simulations predict that we will likely need to synchronise multiple times per hour to ensure that we can have $\gamma$ < 10h, yielding a sufficiently agile subnetwork.

\section{Mathematical Formalism of CLASP}

\begin{enumerate}
    \item Let $\mathcal{N} = \{1, 2, \ldots, N\}$ be the set of miners in the distributed training system, arranged into layers
    \item Each sample $x_k$ follows a pathway $\pi_k \subseteq \mathcal{N}$ through a subset of miners, limited to one miner per layer
    \item Each sample produces a loss $\ell_k \geq 0$
\end{enumerate}

After processing $T$ samples, the orchestrator maintains records:
$$\mathcal{D} = \{(\pi_k, \ell_k) : k = 1, 2, \ldots, T\}$$

where $\pi_k$ is the set of miners that processed sample $k$, and $\ell_k$ is the resulting loss.

For each miner $i \in \mathcal{N}$, compute their \textit{average loss}:

$$\bar{\ell}_i = \frac{1}{|S_i|} \sum_{k \in S_i} \ell_k$$

where $S_i = \{k : i \in \pi_k\}$ is the set of samples that miner $i$ participated in processing.
Given a sufficiently high value of $k$, we can run any statistical outlier detection method of choice to detect malicious miners which have abnormally large contributions to the loss.
\end{document}